\documentclass[11pt]{article}

\newlength{\defbaselineskip}
\usepackage{graphicx}
\usepackage{cite}
\usepackage{url}
\usepackage{algorithmic}
\usepackage{float}

\begin{document}

\title{Dynamic Energy and SLA aware Scheduling of Virtual Machines in Cloud Data Centers}

\title{Energy and SLA aware VM Scheduling}

\author{Radheshyam Nanduri \and Dharmesh Kakadia \and Vasudeva Varma \\
Search and Information Extraction Lab, \\
International Institute of Information Technology, Gachibowli, Hyderabad, India.\\
}

\date{}
\maketitle

\begin{abstract}
With the advancement of Cloud Computing over the past few years, there has been a massive shift from traditional data centers to cloud enabled data centers. The enterprises with cloud data centers are focusing their attention on energy savings through effective utilization of resources. In this work, we propose algorithms which try to minimize the energy consumption in the data center duly maintaining the SLA guarantees. The algorithms try to utilize least number of physical machines in the data center by dynamically rebalancing the physical machines based on their resource utilization. The algorithms also perform an optimal consolidation of virtual machines on a physical machine, minimizing SLA violations. In extensive simulation, our algorithms achieve savings of about 21\% in terms of energy consumption and in terms of maintaining the SLAs, it performs 60\% better than Single Threshold algorithm.

\end{abstract}

\section{Introduction}
\label{sec-introduction}
Virtualization is the technology which enables cloud computing by providing
intelligent abstraction that hides the complexities of underlying software and
hardware. Using this technology, multiple operating system instances called
Virtual Machines (VMs) \cite{virtualization} can be executed on a single
physical machine without interfering each other. Each virtual machine is
installed with its own operating system and acts as an independent machine
running its own applications. The abstraction provided by this technology takes
care of security, isolation of computation and data across the virtual machines
without the knowledge of the user. This gave rise to cloud computing which
commercializes the benefits of consolidation of virtual machines by exposing
them as utility \cite{buyya-vision}. The rise of cloud computing has relieved
many of the enterprises from a massive effort of managing their own data centers
by renting computation resources on-demand from any of the cloud providers.
There are many Infrastructure as a Service (IaaS) providers like Amazon, Rackspace, GoGrid etc., who provide computing power in
\emph{pay-as-you-go} model. These providers provide a simple interface for
managing virtual machine instances on the cloud through web services. Hence we
see more applications being deployed on the cloud framework each day.

\begin{figure*}[ht]
 \centering
\includegraphics[width=.85\textwidth]{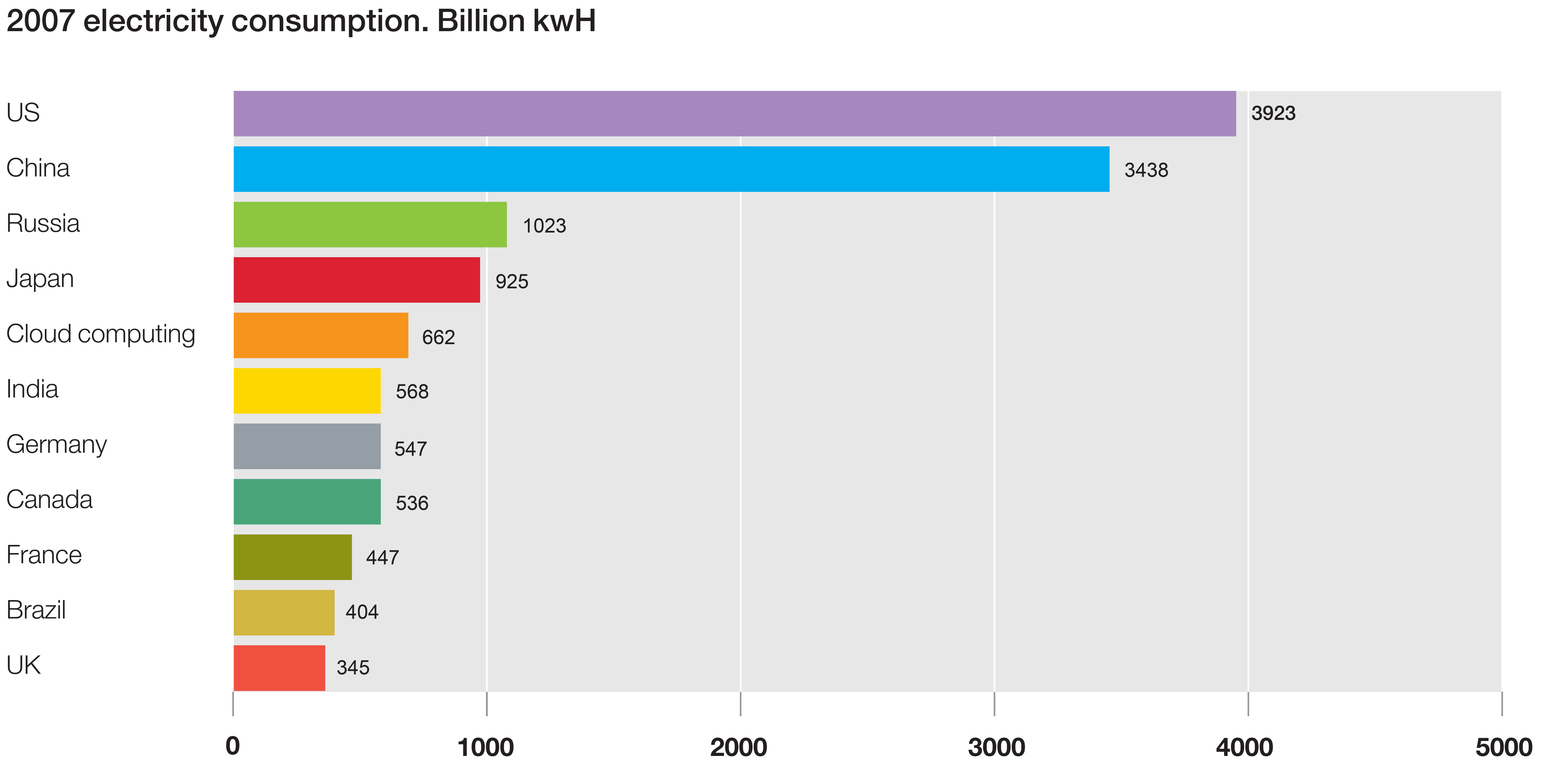}
 \caption{Electricity consumption statistics of various countries in the year
2007. Source: Green Peace International \cite{GreenPeace}.}
 \label{electricityUsed}
\end{figure*}

The cloud providers consolidate the resource requirements of various customers
on to virtual machines across the data center. This inherently does not mean
that these data centers are energy efficient due to consolidation. The cloud
data center administrators have to follow required policies and scheduling
algorithms so as to make their data centers energy efficient. The focus on
\emph{Green Cloud Computing} has been increasing day by day due to shortage of
energy resources. The \emph{U.S. Environmental Protection Agency} (EPA) data
center report \cite{epa} mentions that the energy consumed by data centers has
doubled in the period of 2000 and 2006 and estimates another two fold increase
over the next few years if the servers are not used in an improved operational
scenario. The \emph{Server and Energy Efficiency} Report \cite{1e} states that
more than 15\% of the servers are run without being used actively on a daily
basis. The \emph{Green Peace International} survey \cite{GreenPeace} reports
that the amount of electricity used by Cloud data centers (Figure
\ref{electricityUsed}) could be more than the total electricity consumed by a
big country like India, in the year 2007. This shows that there is a need to
utilize the resources very effectively and in turn save energy.

In this paper, we focus on conserving the energy by effective scheduling and
provisioning of virtual machines without compromising on Service-Level Agreement
(SLA) guarantees. We present scale-up and scale-down algorithms, which try to
consolidate the virtual machines intelligently and reduce the overall energy
usage of the data center.

\subsection{Contributions of this work}

Major contributions of this work are as follows:
  \begin{itemize}
    \item Consolidate the virtual machines effectively based on the resource usage of the virtual machines.
    \item Utilize the physical machines to the maximum extent and put low utilized physical machines to standby mode, by intelligently migrating the load on to other physical machines.
    \item Maintaining the SLA guarantees while effectively saving the power consumed by the data center.
  \end{itemize}

The remainder of the paper is structured as follows. Section \ref{sec-relwork} details about the related work in this area. In Section \ref{sec-prop-algo}, we discuss our allocation, scale-up and scale-down algorithms. Section \ref{sec-eval-results} presents the results of our scheduling algorithms obtained through rigorous simulation and Section \ref{sec-conclusion} concludes the paper with few future directions of this work.

\section{Related Work}	
\label{sec-relwork}

Scheduling has always been a challenging research problem in the field of
computer science. Many scheduling algorithms have been proposed each having its
own pros and cons. 

\subsection{Round Robin, Greedy and Power Save}

Eucalyptus is one of the leading and widely used open source
software packages to set up private cloud infrastructure. Round Robin, Greedy and Power Save algorithms are the virtual machine
scheduling algorithms provided along with it. \emph{Round Robin} algorithm
follows the basic mechanism of allocating the incoming virtual machine requests
on to physical machines in a circular fashion. It is simple and starvation-free
scheduling algorithm which is used in most of the private cloud infrastructures.
The \emph{Greedy} algorithm will allocate the virtual machine to the first
physical machine which has enough resources to satisfy the resources requested
by it. In \emph{Power Save} algorithm, physical machines are put to sleep when
they are not running any virtual machines and are re-awakened when new resources
are requested. First, the algorithm tries to allocate virtual machines on the
physical machines that are running, followed by machines that are asleep.

These algorithms have limited or no support for making scheduling decisions
based on the resource usage statistics. Moreover these algorithms do not take
into account of SLA violations, energy consumed etc., which are very important
factors in real cloud environments. 

\subsection{Dynamic Round Robin}
\label{DRR}

Ching-Chi Lin et. al in \cite{DRR} presented an improved version of Round Robin
algorithm used in Eucalyptus. According to Dynamic Round Robin algorithm, if a
virtual machine has finished its execution and there are still other virtual
machines running on the same physical machine, this physical machine will not
accept any new virtual machine requests. Such physical machines are referred to
as being in `retirement' state, meaning that after the execution of the
remaining virtual machines, this physical machine could be shutdown. And if a
physical machine is in the `retirement' state for a sufficiently long period of
time, the currently running virtual machines are forced to migrate on to other
physical machines and shutdown after the migration operation is finished. This
waiting time threshold is denoted as `retirement threshold'. So, a physical
machine which is in the retirement state beyond this threshold will be forced to
migrate its virtual machines and shutdown. 

Even this algorithm has limited support for making scheduling decisions based on
the resource usage statistics and does not take into account of SLA violations,
energy consumed etc.

\subsection{Single Threshold}

In \cite{SingleThreshold}, the authors propose Single Threshold algorithm which
sorts all the VMs in decreasing order of their current utilization and allocates
each VM to a physical machine that provides the least increase of power
consumption due to this allocation. The algorithm does optimization of the
current allocation of VMs by choosing the VMs to migrate based on CPU
utilization threshold of a particular physical machine called `Single
Threshold'. The idea is to place VMs while keeping the total utilization of CPU
of the physical machine below this threshold. The reason for limiting CPU usage
below the threshold is to avoid SLA violation under a circumstance where there
is a sudden increase in CPU utilization of a VM, which could be compensated with
the reserve. Single Threshold algorithm works better in terms of energy
conservation when compared to Dynamic Round Robin Algorithm discussed in
\ref{DRR}. This algorithm is fairly improved one which takes into consideration
of power consumption and CPU usage of physical machines. 

\subsection{Dynamic Voltage Scaling}

Dynamic Voltage Scaling (DVS) is a power management technique where
under-volting (decreasing the voltage) is done to conserve power and
over-volting (increasing the voltage) is done to increase computing performance.
This technique of DVS has been employed in \cite{buyya-dvs, lee-dvs} to design
power-aware scheduling algorithms that minimize the power consumption. Hsu et
al. \cite{hsu-hpc} apply a variation of DVS called Dynamic Voltage Frequency
Scaling (DVFS) by operating servers at various CPU voltage and frequency levels
to reduce overall power consumption.

\subsection{Dynamic Cluster Reconfiguration}

In \cite{pinheiro, pinheiroxx, nitesh}, the authors proposed systems that
dynamically turn cluster nodes on - to be able to handle the load on the system
efficiently and off - to save power under lower load. The key component of these
algorithms is that the algorithm dynamically takes intelligent re-configuration
decisions of the cluster based on the load imposed on the system. Our work is
mainly inspired by these algorithms which scale the cluster up and down as per
the requirement and save power. We tried to employ the same kind of principle in
a virtualized data center environment. 

In \cite{perez-reconfig}, P\'{e}rez et al. try to achieve a dynamic
reconfiguration using a mathematical formalism, with the use of storage groups
for data-based clusters. A considerable amount of research has been done in the
fields of load balancing and cluster reconfiguration, with prime focus on
harvesting the cycles of idle machines \cite{glunix, douglis, khalidi}. Our work
is based on load balancing and VM migration decisions with prime focus on
reducing the total number of running physical machines.

\section{Proposed Algorithm}
\label{sec-prop-algo}

\begin{figure}[ht]
 \centering
 \includegraphics[width=.75\textwidth]{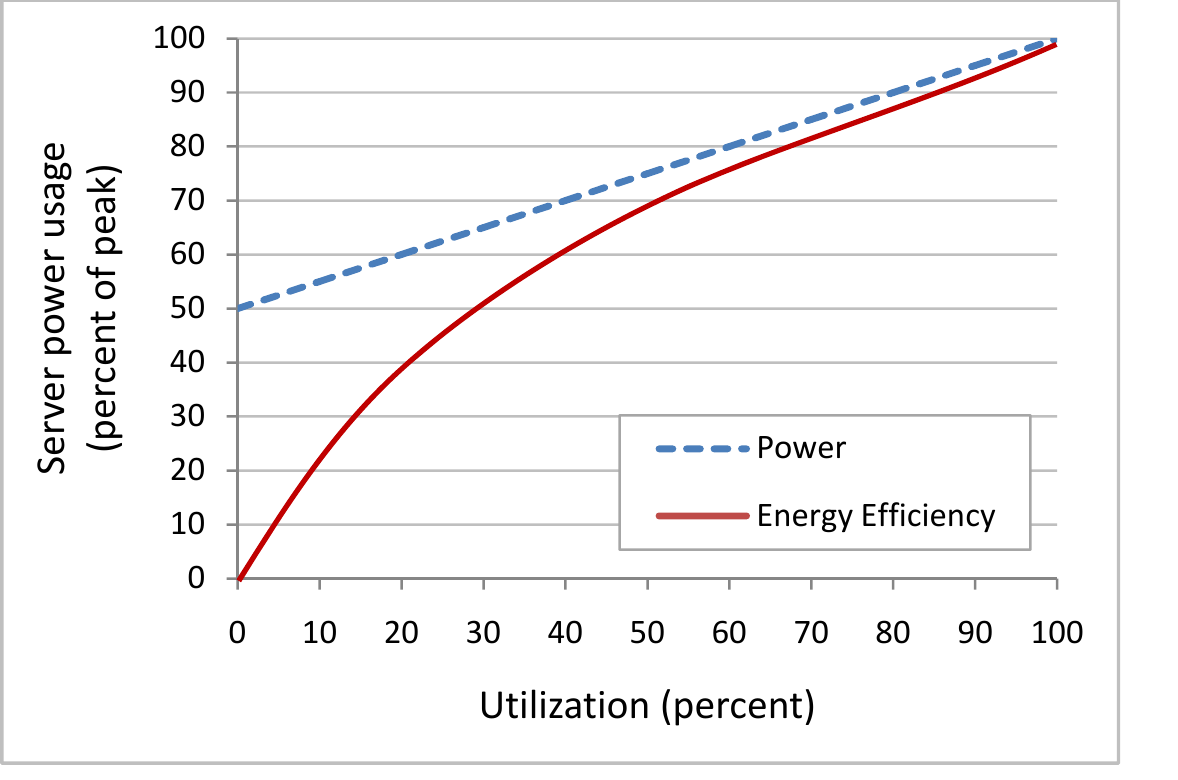}
 \caption{Server power usage and energy efficiency at varying utilization
levels, from idle to peak performance. Even an energy-efficient server still
consumes about half its full power when doing virtually no work. Source:
\cite{energy-prop-comp}.}
 \label{server_utilization}
\end{figure}

Data centers are known to be expensive to operate and they consume huge amounts
of electric power \cite{buyya-vision}. Google's server utilization and energy
consumption study \cite{energy-prop-comp} reports that the energy efficiency
peaks at full utilization and significantly drops as the utilization level
decreases (Figure \ref{server_utilization}). Hence, the power consumption at
zero utilization is still considerably high (around 50\%). Essentially, even an
idle server consumes about half its maximum power. Our algorithms try to
maintain high utilization of physical machines in the data center so as to
utilize energy and resources optimally. In this section, our approach to handle
the scheduling decisions of VMs in the data center is presented.

Initially, we assume that all the physical machines in the data center are put
to standby mode except for few. We start with only one physical machine that is
up and running and only awaken the physical machines as and when required, as
directed by our algorithms discussed ahead. When a new request to allocate a VM
is received by the data center, the request is directed to \emph{Allocation
Algorithm}. The \emph{Allocation Algorithm} takes the decision of allocating the
VM on a particular physical machine.

\subsection{Allocation Algorithm}
\label{sec:allocationalgorithm}

The Allocation Algorithm presented in the Figure \ref{AllocationAlgorithm},
accepts the VM request and tries to fit on to one of the currently running
physical machines. The algorithm tries to fit the virtual machine based on the
resource usage of the target physical machine. The resource usage of the target
physical machine is represented by its \emph{Resource Vector}. Firstly, we
discuss \emph{Resource Vector} which forms the base for our algorithms.

\subsubsection{Resource Vector}

A virtual machine uses the computing resources based on the applications running
on it. Based on the resource usage, a virtual machine can be broadly categorized
as CPU-intensive if it uses high CPU, or memory-intensive if it accounts for more
of memory IO and similarly disk-intensive or network-intensive. But, just identifying this information about a virtual machine does not give its exact resource usage pattern. 
To calculate a better resource usage pattern of a virtual machine, we need to take into account of all the resources used by it, at once. So, we define the resource
usage pattern of a virtual machine as a vector with four components each
denoting CPU, memory, disk and network resources.

  \begin{equation}
  \label{eq:RV}
      Resource Vector (RV) = <E_{cpu}, E_{mem}, E_{disk}, E_{bw}>
  \end{equation} 

where $E_x$ (x is $cpu, mem, disk, bw$) represents the percentage of
corresponding resource used i.e. percentage of total CPU, memory, disk and
network resources used respectively on that physical machine. Since we denote
$E_x$ as percentage of resource used on the physical machine, we represent its
value from 0 to 1.\newline

\textbf{Example: Resource Vector (RV)}

Resource Vector 1 = $<0.70, 0.10, 0.05, 0.05>$ denotes a CPU-intensive vector.

Resource Vector 2 = $<0.70, 0.50, 0.05, 0.05>$ denotes a CPU and memory
intensive vector.  

\subsubsection{Construction of Resource Vector} \label{ConstructionRV}

The resources used by a virtual machine are logged at regular intervals at the
hypervisor level. Resource Vector (RV) of virtual machine is represented as
$RV_{vm}$. $E_x$ in $RV_{vm}$ of the virtual machine is calculated by averaging
its corresponding resource usage (say $E_{cpu}$) over a period of time $\Delta$
(previous $\Delta$ time units in the history). For example, $E_{cpu}$ at any
time $\tau$ is the average percentage utilization of CPU by the virtual machine
between $\tau - \Delta$ and $\tau$. \newline

\textbf{Handling Resource Vector in Heterogeneous Environment:}\emph{Resource Vector} of a VM ($RV_{vm}$) is the vector representation of
percentage of resources utilized by the VM on a physical machine. But since the
data center could be heterogeneous, this $RV_{vm}$ may not be uniform across
different physical machines because of diverse resource capacities. To handle
such heterogeneous data center environments, the resource vector could be
modified as $RV_{vm}(PM)$, denoting resource vector of a VM on a particular PM.\newline

\textbf{Example RV in heterogeneous environment:} Resource Vector RV of a VM on a physical machine $PM_1$ is given as follows:

\begin{equation}
	RV_{vm}(PM_1) = <E_{cpu}, E_{mem}, E_{disk}, E_{bw}>
\end{equation}

similar to Equation \ref{eq:RV}, where

\begin{equation}
	E_{cpu} = \frac{CPU\ used\ by\ VM}{max\ CPU\ capacity\ of\ PM_1}
\end{equation}

Similarly, the rest of the components of $RV_{vm}(PM_1)$, which are $E_{mem}$,
$E_{disk}$, $E_{bw}$ can be calculated.

So, given the resource vector of a VM on a physical machine say, $PM_1$ i.e.,
$RV_{vm}(PM_1)$, we can calculate its resource vector corresponding to another
physical machine say, $PM_2$ denoted by $RV_{vm}(PM_2)$. The calculation is
straight forward as the information about resource capacities of both the
physical machines is available to the system.

Next, the Allocation algorithm tries to allocate the new VM request on to the
physical machine on which it fits the best. To check whether a VM perfectly fits
on a running physical machine, we follow \emph{Cosine Similarity model}.

\subsubsection{Cosine Similarity Model}

Cosine similarity gives the measure of the angle between two vectors. If the angle between two vectors is small, then they are said to possess similar alignment. The cosine of two vectors lies between -1 and 1. If the vectors point in the same direction, the cosine between them is 1 and the value decreases and falls to -1 with an increase in angle between them.

Using Euclidean dot product, the cosine of two vectors is defined as

  \begin{equation}  
    \mathbf{a}\cdot\mathbf{b}=\left\|\mathbf{a}\right\|\left\|\mathbf{b}\right\|\cos\theta
  \end{equation}

And the similarity is shown as follows,

  \begin{equation} \label{eq:CosineSimilarity}
      \textbf{similarity} = {A \cdot B \over \|A\| \|B\|} = \frac{ \sum_{i=1}^{n}{A_i \times B_i} }{ \sqrt{\sum_{i=1}^{n}{(A_i)^2}} \times \sqrt{\sum_{i=1}^{n}{(B_i)^2}} }
  \end{equation}

The Allocation Algorithm uses this similarity model and tries to allocate the
incoming virtual machine request on to a physical machine based on the
\textbf{similarity} measure between $RV_{vm}(PM)$ of incoming VM and RV of
physical machine (denoted by $RV_{PM}$ which will be discussed later). 

This idea of similarity is used to allocate dissimilar VMs on a physical
machine. By similar/dissimilar VMs, we are referring to the
similarity/dissimilarity in resource usage patterns of the VMs. For example, if
VM1 is CPU-intensive, we would not want VM2 which is also CPU-intensive, to be
allocated on same physical machine since there may be a race condition for CPU
resource. By allocating dissimilar VMs on the physical machine, following
benefits could be achieved.

  \begin{enumerate}
    \item Race condition for the resources between the VMs could be minimized.
    \item The physical machine would be trying to use all the resources, increasing its overall utilization.
  \end{enumerate}

\textbf{Reason for choosing Cosine Similarity model:} The reason for choosing Cosine Similarity model over the other similarity models
is that, it is simpler and takes into consideration of similarity measure of
each component of the vector. And this perfectly suits our requirement of
comparing usage patterns of different resources at a time.

Before moving forward, we shall discuss about Resource Vector of a physical
machine, $RV_{PM}$. $RV_{PM}$ is the percentage of resources used on the
physical machine. It is similar to $RV_{vm}(PM)$ denoting the percentage of
resources used on the physical machine i.e., the usage accounted due to sum of
the resources consumed by all the virtual machines running on that particular
physical machine. $RV_{PM}$ can be shown as follows,

  \begin{equation}
      RV_{PM} = <E_{cpu\_used}, E_{mem\_used}, E_{disk\_used}, E_{bw\_used}>
  \end{equation}

where $E_{x\_used}$ (x is $cpu, mem, disk, bw$) represents the percentage of
corresponding resource used i.e. percentage of total CPU, memory, disk and
network resources used respectively on the physical machine. Similarly,
$PM_{free}$ is resource vector which represents the free resources available on
physical machine.

\subsubsection{Calculation of Similarity}

As discussed, Allocation Algorithm uses the cosine similarity measure to find a
physical machine that is most suitable for the incoming VM request. To use the
cosine similarity model, we need to know the RV of the incoming VM. But, since
the incoming VM's resource usage may not be known ahead of it's allocation, we
make an initial assumption to take a default $RV_{vm}(PM)$. The default
$RV_{vm}(PM)$ is assumed to be $<0.25, 0.25, 0.25, 0.25>$. Once the VM is
allocated and run for a time period of $\Delta$, its exact $RV_{vm}(PM)$ could
be found by the mechanism discussed in \ref{ConstructionRV}. 

To avoid race condition for resources between the VMs, we need to allocate VMs
of dissimilar nature. We propose two different methods of calculating similarity
measure which are based on Cosine Similarity. \newline

\textbf{Method 1 - Based on dissimilarity: }
In this method, we calculate the cosine similarity between RV of the incoming VM
and $RV_{PM}$. And, we select a running physical machine which gives
\emph{least} cosine \textbf{similarity} measure with the incoming VM. The
\emph{least} cosine similarity value implies that the incoming VM is mostly
dissimilar to the physical machine in terms of resource usage patterns.

By equation \ref{eq:CosineSimilarity} we arrive at following formula,

  \begin{equation}
      \textbf{similarity} = {RV_{vm}(PM) \cdot RV_{PM} \over \|RV_{vm}(PM)\|
\|RV_{PM}\|}
  \end{equation}\newline

\textbf{Method 2 - Based on similarity: }
In this method, we calculate the cosine similarity between RV of the incoming VM
and $PM_{free}$. 

  \begin{equation}
      \textbf{similarity} = {RV_{vm}(PM) \cdot PM_{free} \over \|RV_{vm}(PM)\| \|PM_{free}\|}
  \end{equation}\newline

We select a running physical machine which gives \emph{maximum} cosine
\textbf{similarity} measure with the incoming VM. The \emph{maximum} cosine
similarity value implies that the incoming VM's resource requirements are most
compatible with the free resources of physical machine. 

The \textbf{similarity} value lies between 0 and 1 since we are not dealing with
negative physical resource values.\newline

\textbf{Difference between Method 1 and 2:}
The similarity methods discussed above help in consolidating VMs on a physical
machine without a race condition for resources. There is a subtle difference
between the proposed methods. Method 1 tries to allocate VMs which are
dissimilar in resource usage patterns. This method helps in achieving the
consolidation of VMs with diverse resource usage patterns. While, Method 2 tries
to allocate a VM which could properly consume the underutilized resources of the
physical machine. This method inherently makes sure that race condition for
resources is avoided and at the same time improves the utilization of the
physical machine.

Before discussing further algorithms, we present the utilization model for a
physical machine upon which the following algorithms are based on.

\subsubsection{Utilization model}

Our work considers multiple resources viz. CPU, memory, disk and network of a
physical machine. It is difficult to incorporate utilizations of each of the
resources individually into the algorithms. Hence, we come up with a unified
model that tries to represent the utilization of all these resources into a
single measure, $U$. The unified utilization measure, $U$ is considered to be a
weighted linear combination of utilizations of individual resources. It is given
as follows,

  \begin{equation}
      U = \alpha \times E_{cpu} + \beta \times E_{mem} + \gamma \times E_{disk} + \delta \times E_{bw}
  \end{equation}

where, $\alpha, \beta, \gamma, \delta \in [0, 1] $ can be weighed accordingly by
the the administrator as per the requirements. And,

  \begin{equation}
      \alpha + \beta + \gamma + \delta = 1
  \end{equation}

So we try to measure the utilization of any physical machine or virtual machine
through a single parameter, $U$. This unified single parameter, $U$ is
introduced for simplicity reasons which reduces the difficulty in taking into
consideration of multiple parameters into our algorithms.

The Allocation Algorithm not only tries to consolidate dissimilar VMs but also
makes sure that the physical machine is not overloaded after the allocation of
VM. Hence, first the similarity measure between the VM and the physical machine
is calculated. If the algorithm finds that the similarity measure is good enough
to accommodate the VM on the physical machine we proceed to next step. In the
next step, the algorithm calculates the estimated $U$ after the VM allocation on
the target physical machine (the machine which is suggested by similarity
measure) ahead of its actual allocation. And the VM allocation is considered
only if $U$ after allocation, i.e., the estimated utilization of the physical
machine after the allocation of VM on it, is less than the value ($U_{up}$ $-$
\emph{buffer}).

If $U$ after allocation, is greater than the value ($U_{up}$ $-$ \emph{buffer}), we do not consider that physical machine as the allocation may overload it. Instead we take the physical machine which is next best in terms of similarity measure and find its $U$ after allocation. The physical machine is accepted if $U$ after allocation is less than the value ($U_{up}$ $-$ \emph{buffer}), else we repeat the same procedure by taking the physical machine with next best similarity measure. The details of this value ($U_{up}$ $-$ \emph{buffer}) is
discussed clearly later.

\renewcommand{\algorithmicforall}{\textbf{for each}}

\begin{figure}[H]
\begin{algorithmic}[1]
	\STATE \emph{\textbf{Allocation Algorithm}}(VMs to be allocated)
	    \item[] \COMMENT{`VMs to be allocated' is the argument passed to this algorithm}
	\FORALL{VM $\in$ VMs to be allocated}
		\FORALL{PM $\in$ Running PMs}
			\item[] \COMMENT{physical machine is represented as PM}
			\STATE $similarity_{PM} = calculateSimilarity(RV_{vm}(PM), RV_{PM})$
			\item[] \COMMENT{similarity is calculated using any of the two methods discussed}
			\STATE  add $similarity_{PM}$ to $queue$
		\ENDFOR
		\item[]
		\STATE sort $queue$ in ascending values of $similarity_{PM}$ \COMMENT {if Method 1 is used} $or$ \\
		 sort $queue$ in descending values of $similarity_{PM}$ \COMMENT {if Method 2 is used}
		\item[]
		\FORALL{$similarity_{PM}$ in $queue$}
			\STATE $target_{PM}$ = PM corresponding to $similarity_{PM}$

			\IF{$U$ after allocation on $target$ PM $<$ ($U_{up}$ $-$ \emph{buffer})}
				\STATE $allocate($VM, $target$ PM)
				\item[] \COMMENT{VM is allocated on $target$ PM}
				\RETURN SUCCESS
			\ENDIF
		\ENDFOR
		\item[]
		\RETURN FAILURE \COMMENT{VM can't be allocated on any of the running machines}
	\ENDFOR
\end{algorithmic}
\caption{Allocation Algorithm. The VMs are consolidated on physical machines
based on similarity measure.}
\label{AllocationAlgorithm}
\end{figure}

If the algorithm fails to find any running physical machine which satisfies both
the above conditions, then it awakens one of the standby physical machines and
allocates the VM on it. The calculation of estimated $U$ after allocation is
straight-forward since we have enough information about the resource vectors of
VMs and physical machines. 

After the allocation, the resource usage of each physical machines is monitored
at regular intervals. And if the utilization of a physical machine reaches an
administrator specified threshold (\emph{Scale-up Threshold}, $U_{up}$), we
follow the following Scale-up Algorithm.

\subsection{Scale-up Algorithm}

If the utilization, $U$ of any physical machine is observed to be greater than
$U_{up}$ for a consistent time period $T$, the Scale-up Algorithm is triggered.
The Scale-up Algorithm presented in Figure \ref{Scale-up Algorithm}, then tries
to bring down $U$ of the physical machine by migrating the VM with
\textbf{\emph{highest utilization}} on that particular physical machine to
another physical machine. Firstly, the Scale-up algorithm hands over the VM with
high utilization on that overloaded physical machine to Allocation Algorithm for
suitable migration. Then, the Allocation Algorithm tries to consolidate that
particular VM on any of the other already running physical machines, duly taking
into consideration that the migration does not overload the target physical
machine as well. If the Allocation Algorithm succeeds in finding a physical
machine to allocate the VM, the migration of the VM is instantiated on to the
target physical machine. But, if the Allocation Algorithm fails to find a
suitable physical machine, then one of the standby physical machines is awakened
and migration of the VM is instantiated on to it. By doing this we bring down
the $U$ of the physical machine below $U_{up}$.

Addition of standby physical machines to the running physical machines happens
only when required, to handle the rise in resource requirement. This makes sure
that the physical machines are used very optimally, conserving a lot of energy.

\begin{figure}[h]
\begin{algorithmic}[1]
	\STATE \emph{\textbf{Scale up Algorithm}}()
	\IF{$U > U_{up}$}
	  \item[] \COMMENT{if $U$ of a PM is greater than $U_{up}$}
	  \STATE VM = VM with $max$ $U$ on that PM
	  \STATE Allocation Algorithm(VM)
	\ENDIF
	\IF{Allocation Algorithm fails to allocate VM}
		\STATE $target$ PM = add a standby machine to running machine
		\STATE $allocate($VM, $target$ PM) 
	\ENDIF
\end{algorithmic}
\caption{Scale-up Algorithm. Upon reaching the scale-up trigger condition, the
above algorithm is executed.}
\label{Scale-up Algorithm}
\end{figure}

Similarly, if the utilization of a physical machine goes down below an
administrator specified threshold (\emph{Scale-down Threshold}, $U_{down}$), we
follow the following Scale-down Algorithm.

\subsection{Scale-down Algorithm}

If the utilization, $U$ of any physical machine is observed to be lower than
$U_{down}$ for a consistent time period $T$, the Scale-down Algorithm is
triggered. This suggests that the physical machine is under-utilized. So, the
Scale-down Algorithm presented in Figure \ref{Scale-down Algorithm}, tries to
migrate VMs on that particular under-utilized physical machine to other running
physical machines and put it to on standby mode. The VMs on the physical machine
are handed over to Allocation Algorithm one after the other for allocation on
any other running physical machines, duly taking into consideration that the
target physical machines are not overloaded. If the Allocation Algorithm
succeeds in finding suitable physical machine where it can consolidate these
VMs, the migration of such VMs is initiated on to the target physical machine.
The physical machine is then put in standby mode after all the migration
operations are performed. But, if the Allocation Algorithm fails to find a
suitable physical machine, the VMs are allowed to run on the same physical
machine. \newline

\begin{figure}[h]
\begin{algorithmic}[1]
	\STATE \emph{\textbf{Scale down Algorithm}}()
	\IF{$U < U_{down}$}
	  \item[] \COMMENT{if $U$ of a PM is less than $U_{down}$}
	  \STATE Allocation Algorithm(VMs on PM)
	\ENDIF
\end{algorithmic}
\caption{Scale-down Algorithm. Upon reaching the scale-down trigger condition,
the above algorithm is executed.}
\label{Scale-down Algorithm}
\end{figure}

\textbf{Reason for using Threshold:}
Scale-up and Scale-down algorithms are triggered when it is observed that $U$ of
a physical machine is above or below $U_{up}$, $U_{down}$ thresholds for a
consistent period of time respectively. These thresholds make sure that the
physical machines are neither over-loaded nor under-utilized. Preventing the
utilization of physical machine above $U_{up}$ helps in reserving sufficient
resources for any sudden surge in utilizations of any of the VMs. This reserve
compute resources greatly helps in avoiding any SLA violations. Similarly, usage
of $U_{down}$ helps in conserving energy by putting an under-utilized machine to
standby mode. To trigger these algorithms, it is a necessary condition that the
utilization activity on physical machine should persist consistently for a
certain period of time. Sometimes there could be a sudden surge in utilization
of a VM and may just persist for small duration. By imposing this condition we
could avoid unnecessary migration of VMs during such conditions.

Since these thresholds are percentage utilizations of physical machines, the
algorithms work unchanged for heterogeneous data centers.\newline

\textbf{Reason for using buffer:}
Before a VM is allocated on a physical machine using Allocation Algorithm, its
utilization $U$ after the VM's allocation is calculated upfront. And the
allocation of that VM is considered only if $U$ after the allocation is less
than ($U_{up}$ - \emph{buffer}). This \emph{buffer} value is considered to make
sure that the utilization does not reach $U_{up}$ immediately after allocation,
which avoids scale-up operation.\newline

\textbf{Selection of standby physical machines while scaling up:}
During the scale-up operation, a standby physical machine may be re-awakened to
accommodate VMs. The machine which is least recently used is picked up while
selecting a standby physical machine. This makes sure that all the machines in
the data center are uniformly used and avoids hot-spots.\newline

\textbf{Difference between Scale-up and Scale-down Threshold:}
$U_{up}$ and $U_{down}$ are set by the administrator of the data center as per
the requirements. Difference in these values should be made sufficiently large
so that the data center does not experience a jitter effect of scaling up and
down very frequently.

\section{Evaluation and Results}
\label{sec-eval-results}

The cloud data center architecture is simulated and the results are generated
over it. The simulator is written in Java. 

\subsection{Simulation Model}

Our simulator simulates the cloud data center from a granularity level of
physical machines, virtual machines running on it, to applications running on
each virtual machine. Each physical machine could be designed with its own
resource specification. Each virtual machine could be assigned to any physical
machine dynamically with requested amount of resources. One or many applications
could be run on each virtual machine with its own resource requirement dynamics.
The simulator has the provision to incorporate scheduling algorithms which guide
the allocation of resources in the data center. The simulator takes care of the
amount of energy consumed using the model discussed in Google's server
utilization and energy consumption study \cite{energy-prop-comp}. The simulator
is designed with the following SLA model. \newline

\textbf{SLA Model:} An SLA violation is considered at the process scheduling
level of hypervisor, whenever any requested resource could not be met to any
virtual machine. In simpler terms, during the scheduling of VMs on a physical
machine by the hypervisor (scheduling the VMs is a kind of process scheduling in
the operating system), a violation of SLA is considered, whenever requested
resources such as the amount of CPU, memory, disk or network could not be
supplied to any virtual machine.

\subsection{Experimental Set-up and Dataset}

The simulation is performed on a machine with Intel core 2 Duo, 2.4 GHz
processor with 2 GB of memory and 500 GB of hard disk which runs Ubuntu 10.04
LTS (Lucid Lynx). 

Rigorous simulation is carried out with various distinctive workloads, based on
the real life data center usage, as the input dataset to the simulator. The
simulator and algorithm parameters are specified in Table \ref{tbl-sim-params}.
To verify the efficacy of our algorithms, we compared them to Single Threshold
algorithm and the results are recorded as follows.

\begin{table}
\renewcommand{\arraystretch}{1.3}
\caption{Simulation and Algorithm Parameters}
\label{tbl-sim-params}
\centering
\begin{tabular}{p{3.5cm}|p{4cm}}
\hline
\bfseries Parameter & \bfseries Value\\
\hline\hline
Scale-up Threshold, $U_{up}$ & [0.25, 1.0]\\
\hline
Scale-down Threshold, $U_{down}$ & [0.0 to 0.4]\\
\hline
\emph{buffer} & [0.05 to 0.5]\\
\hline
Similarity Threshold & [0, 1]\\
\hline
Similarity Method & Method 1 or 2\\
\hline
Number of physical machines & 100\\
\hline
Specifications of physical machines & Heterogeneous\\
\hline
Time period for which resource usage of VM is logged for exact $RV_{vm}$
calculation, $\Delta$ & 5 minutes\\
\hline
\end{tabular}
\end{table}

\subsection{Energy Savings}

\subsubsection{Effect of Scale up Threshold}

\begin{figure}
 \centering
 \includegraphics[width=.75\textwidth]{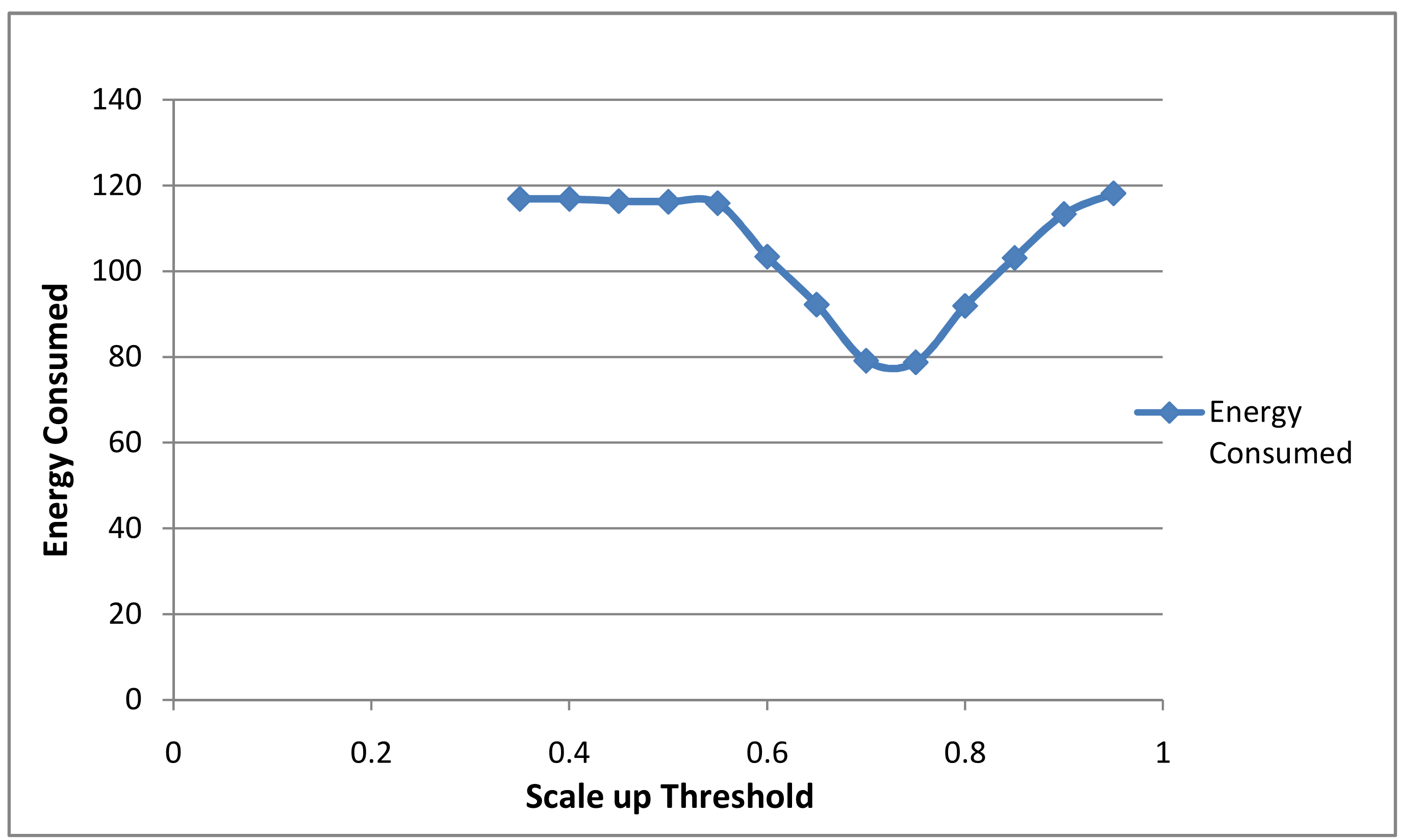}
 \caption{The graph demonstrates the effect of Scale up Threshold on energy
consumption (in kWh). We see a sudden drop of energy consumption when $U_{up}$
is around 0.70 to 0.80.}
 \label{energy_consumed_with_ScaleupThreshold}
\end{figure}

Experiments are carried out on our algorithms to find out the effect on energy
consumption for various values of $U_{up}$ and the output is plotted in Figure
\ref{energy_consumed_with_ScaleupThreshold}. The curve shows a dip when $U_{up}$
is around 0.70 to 0.80 indicating a sudden drop in the energy consumed. 

The curve says that the $U_{up}$ should not be too high or too low and its
optimal value is around 0.70 to 0.80. If $U_{up}$ is low, Scale-up algorithm
tries to run more physical machines to accommodate VMs. And when $U_{up}$ is too
high, we see more number of VMs getting consolidated in the machine and few
surges in the usage of VMs could lead to running new physical machines. Hence,
we see a gradual increase in the energy consumption after 0.80. 

\subsubsection{Effect of scaling down}

\begin{figure}
 \centering
 \includegraphics[width=.75\textwidth]{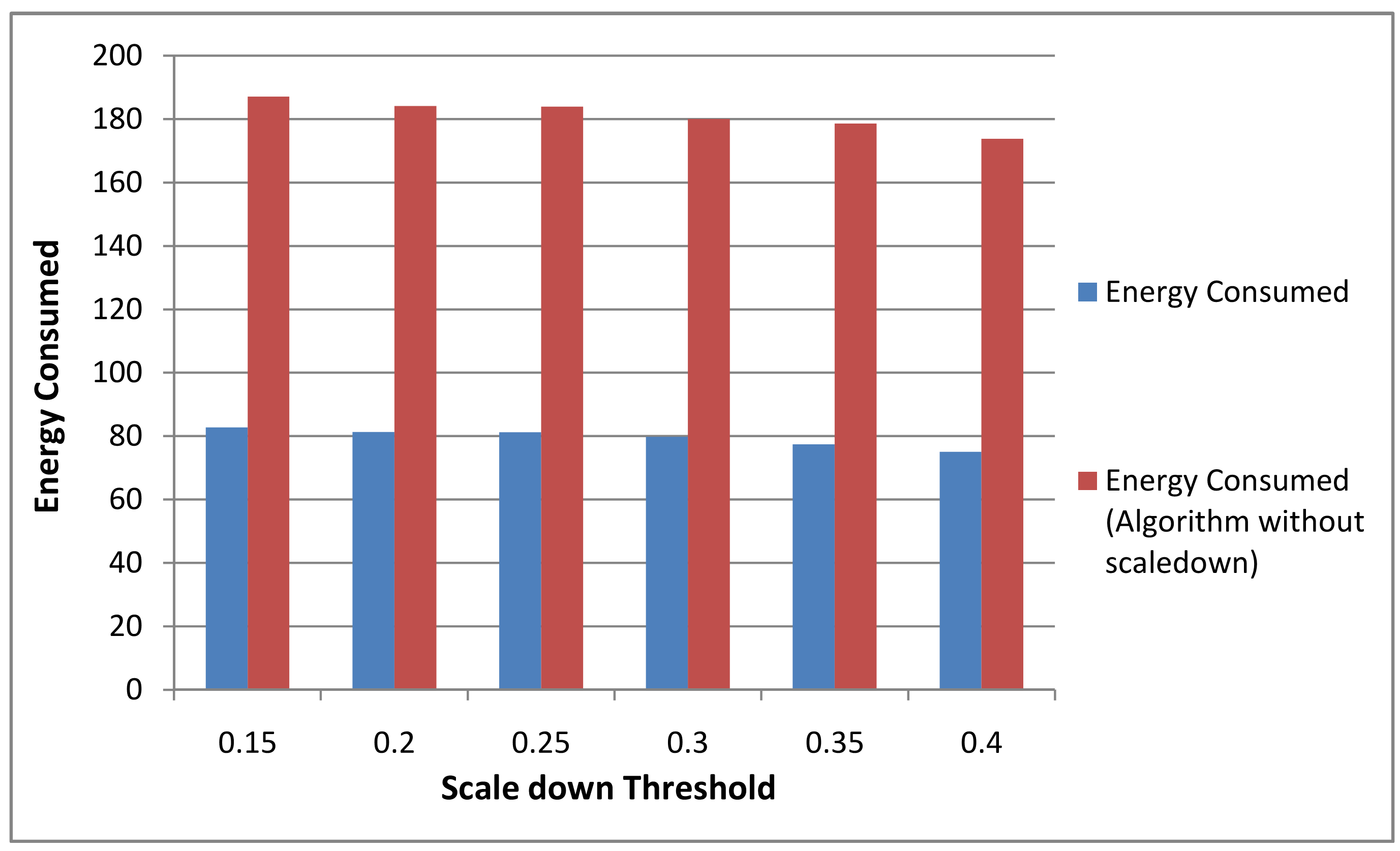}
 \caption{The graph demonstrates the effect of Scale down Threshold on energy
consumption (in kWh). Algorithm with scale down procedure enabled, performs
better in terms of energy conservation.}
 \label{energy_consumed_with_ScaledownThreshold}
\end{figure}

Figure \ref{energy_consumed_with_ScaledownThreshold} demonstrates the use of
having a threshold to put machines to sleep and its effect on energy
conservation.

The graph shows that the energy consumed by our algorithms with scale down
algorithm enabled, is much lower than the algorithm without scale down
procedure. Scaling down of machines when there is not enough load on them could
directly save upto 50\% of energy as demonstrated in the figure. Higher the
value of $U_{down}$, more the physical machines that are scaled down. At the
same time, $U_{down}$ should not be too high, which could result in a jitter
effect of scaling up and down, due to a low difference between $U_{up}$ and
$U_{down}$, which was discussed earlier.

\subsection{SLA violations}

\subsubsection{Effect of Similarity Threshold}

\begin{figure}
 \centering
 \includegraphics[width=.75\textwidth]{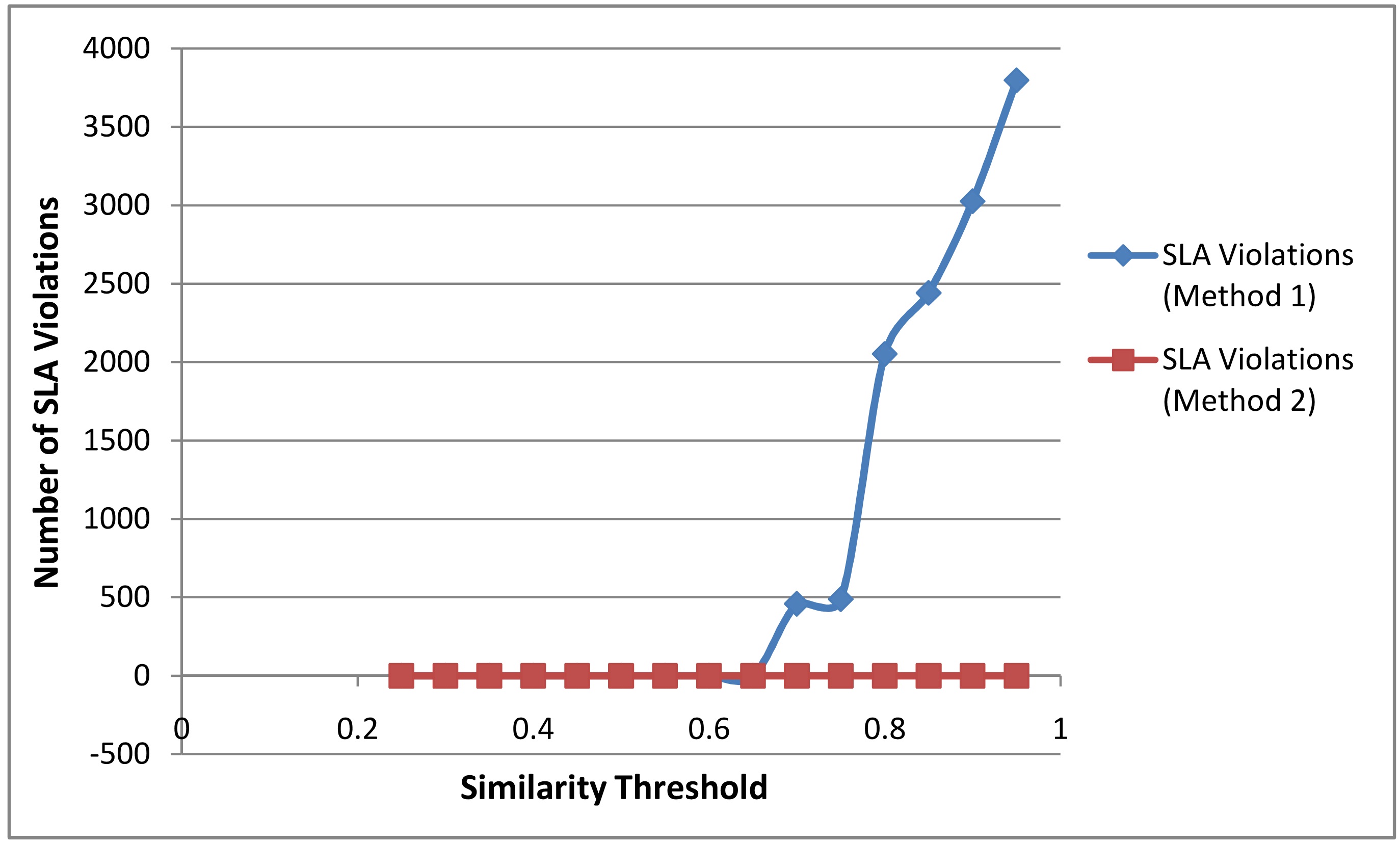}
 \caption{The graph demonstrates the effect of Similarity Threshold on number of
SLA violations. Method 2 performs very well with zero violations for any
Similarity Threshold.}
 \label{SLA_Violation_with_SimilarityThreshold}
\end{figure}

In Figure \ref{SLA_Violation_with_SimilarityThreshold} we try to compare our
results with Method 1 and Method 2 similarity measures. The \textbf{similarity}
value lies between 0 and 1 since we are not dealing with negative resource
values. Method 1 works well with zero violations for lower values of
\emph{Similarity Threshold} as expected, i.e. for values less than 0.6, since
dissimilar VMs are perfectly consolidated with lower threshold values. We
observe that Method 2 works even better in terms of SLA violations which shows
no SLA violations for any \emph{Similarity Threshold}. This is because Method 2
takes into consideration of available resources in the first place, even before
performing consolidation. This proves as an advantage in case of Method 2 and
hence Method 2 is better than Method 1 in terms of consolidation.

\subsubsection{Effect of Scale up Threshold}

\begin{figure}[h]
 \centering
 \includegraphics[width=.75\textwidth]{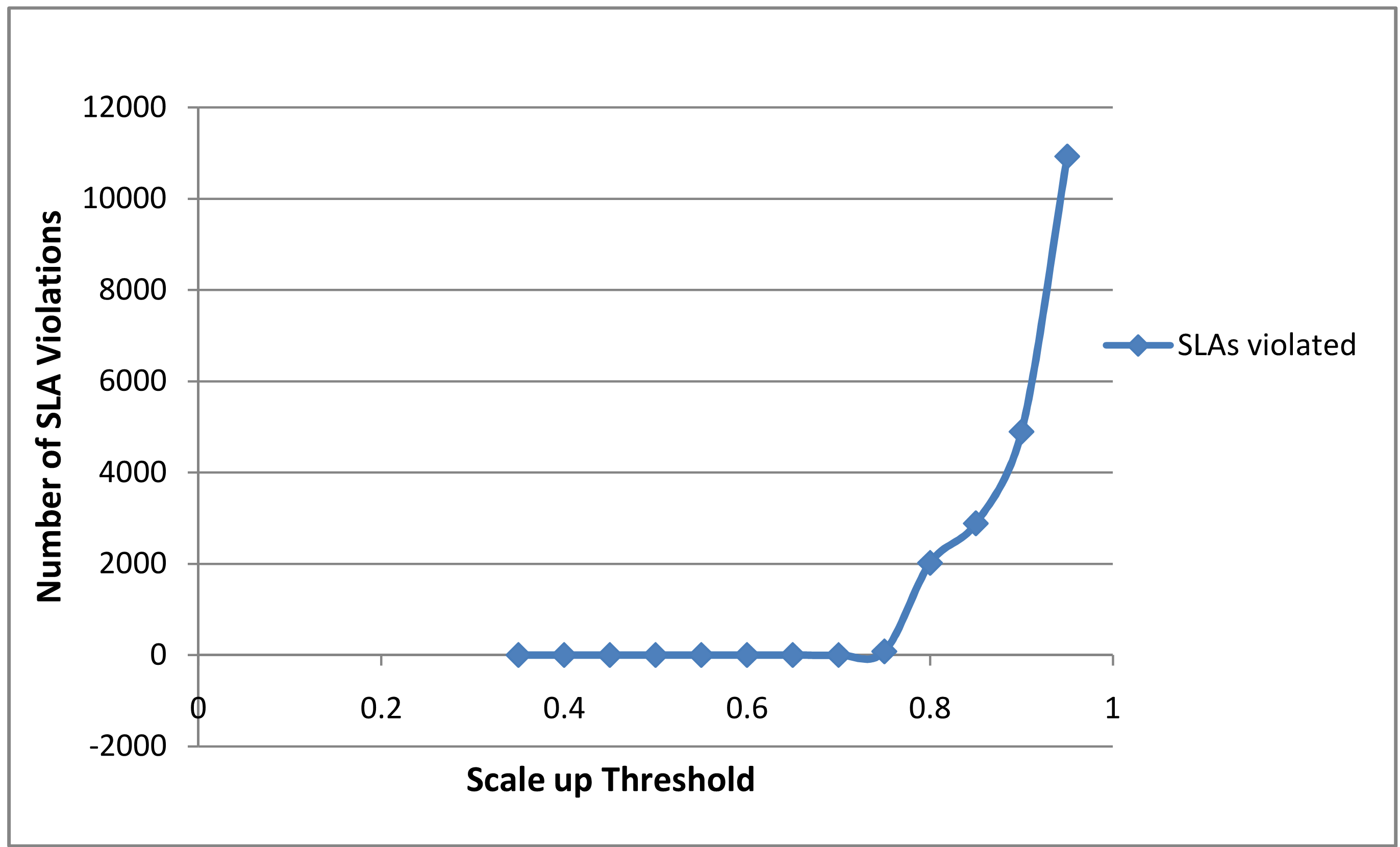}
 \caption{The graph demonstrates the effect of Scale up Threshold on number of
SLA violations. No violations occur for lower values of Scale up Threshold.}
 \label{SLA_Violation_with_ScaleupThreshold}
\end{figure}

In Figure \ref{SLA_Violation_with_ScaleupThreshold} we try to compare the effect
of $U_{up}$ on the number of SLA violations. A very highly stochastic workload
is imposed to the simulator to test this experiment. We see that the SLAs are
not violated for lower $U_{up}$ values. But as $U_{up}$ increases, more VMs get
consolidated on a single physical machine. And when there is a sudden surge in
usage of few of the VMs on this machine, there is not enough free resources to
handle the immediate requirement, which leads to SLA violations. 

\subsection{Effect of buffer}

\begin{figure}[h]
 \centering
 \includegraphics[width=.75\textwidth]{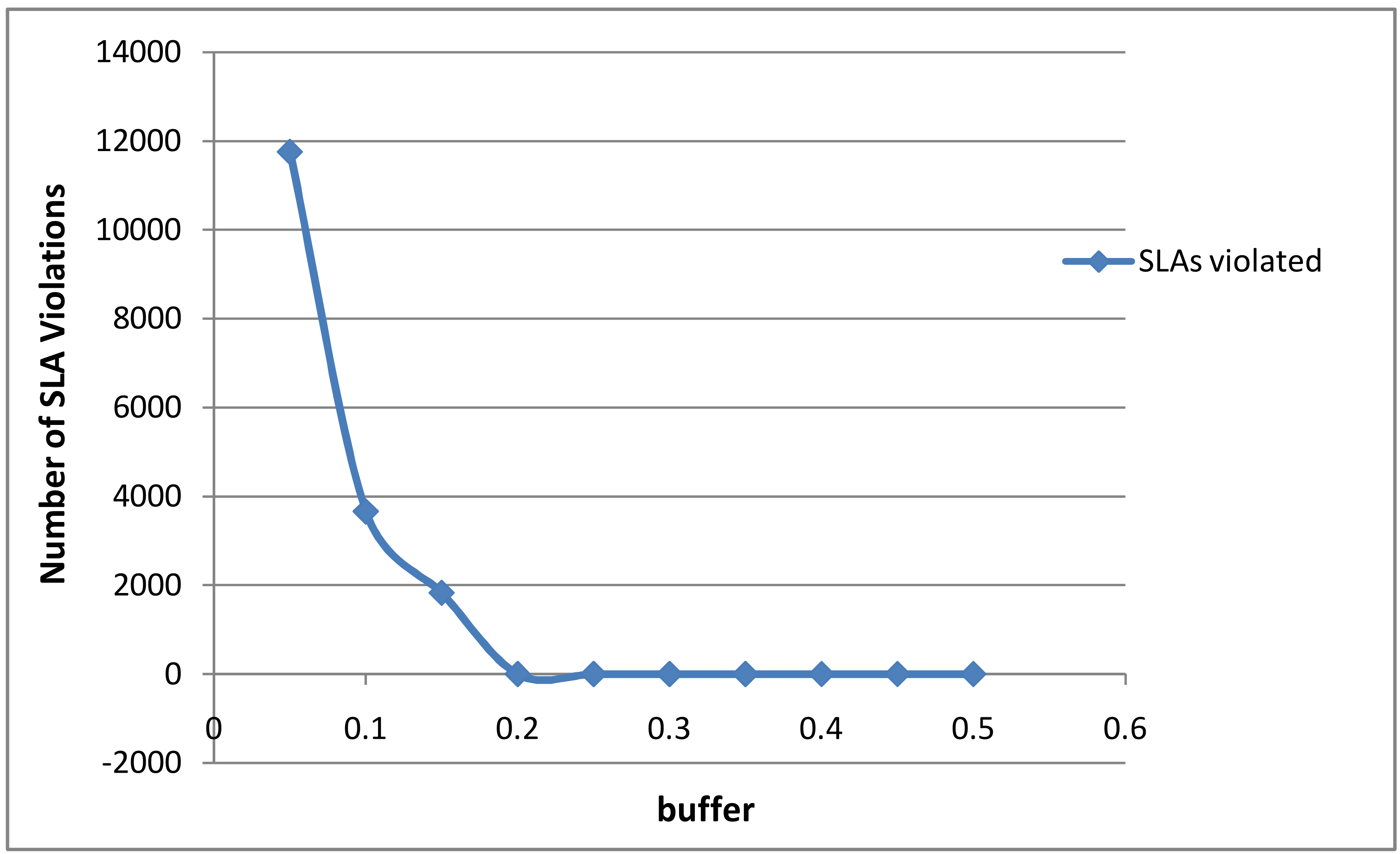}
 \caption{The graph demonstrates the effect of buffer on SLA violations. The
number of SLA violations drop to zero with a buffer value of more than or equal
to 0.2.}
 \label{SLA_Violation_with_buffer}
\end{figure}

\begin{figure}[h]
 \centering
 \includegraphics[width=.75\textwidth]{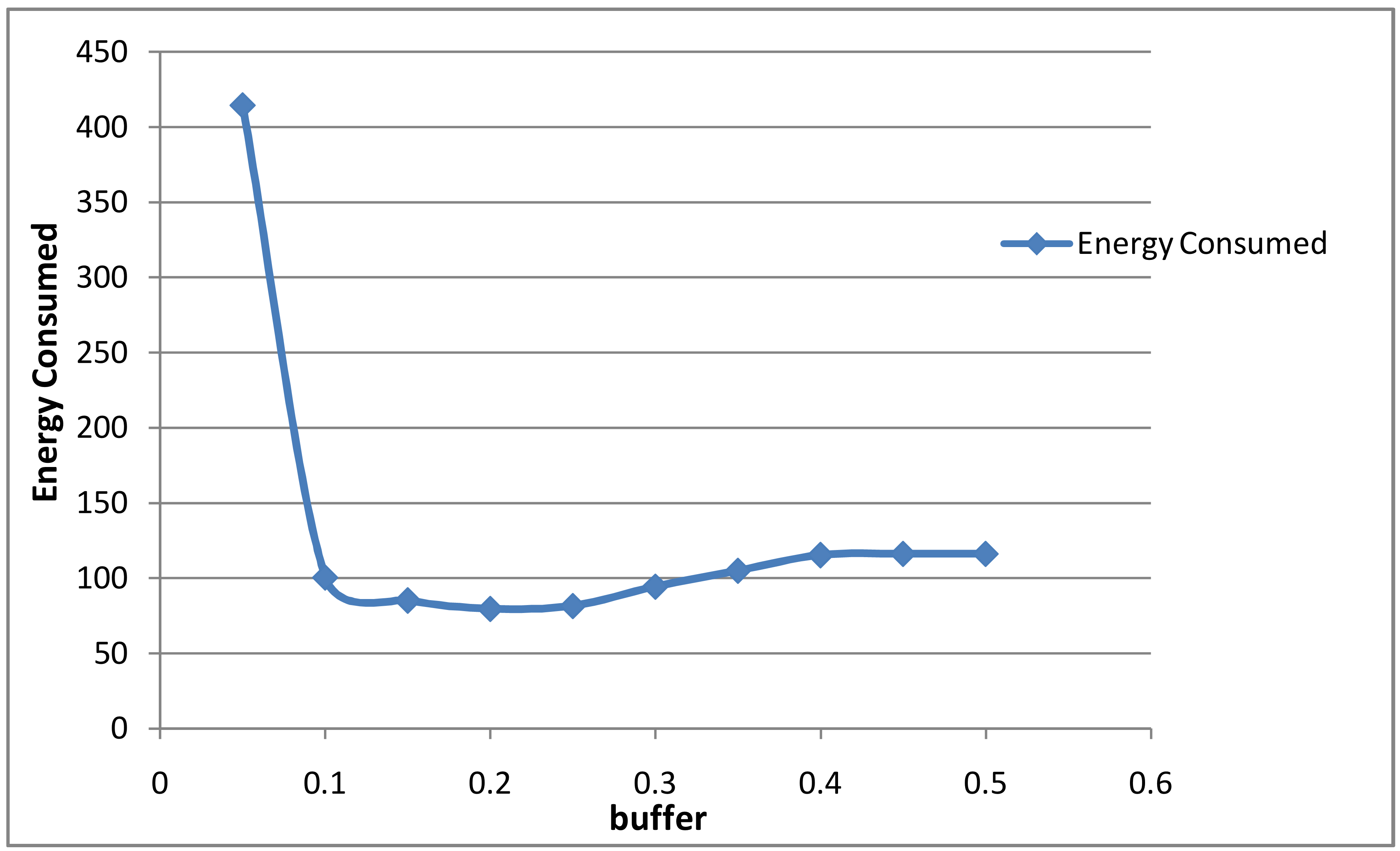}
 \caption{The graph demonstrates the effect of buffer on energy consumption (in
kWh). We see a sudden drop of energy consumption when buffer is around 0.20, but
steadily increases beyond it.}
 \label{energy_consumed_with_buffer}
\end{figure}

An additional padding called as \emph{buffer} is provided to Allocation
Algorithm shown in Figure \ref{AllocationAlgorithm}, to avoid SLA violations.
The Figure \ref{SLA_Violation_with_buffer} shows the advantage of having
\emph{buffer} on SLA violations.We see in the curve that as \emph{buffer}
increases, the number of SLA violations drop to zero, which is as expected. The
\emph{buffer} value has to be used very economically in conjunction with
$U_{up}$ and optimal value is around 0.2. Increase in \emph{buffer} creates more
hindrance to consolidation, causing a steady increase in energy consumption
which is shown in Figure \ref{energy_consumed_with_buffer}.

\subsection{Effectiveness of our algorithm against Single Threshold Algorithm}

\begin{figure}[H]
 \centering
 \includegraphics[width=.75\textwidth]{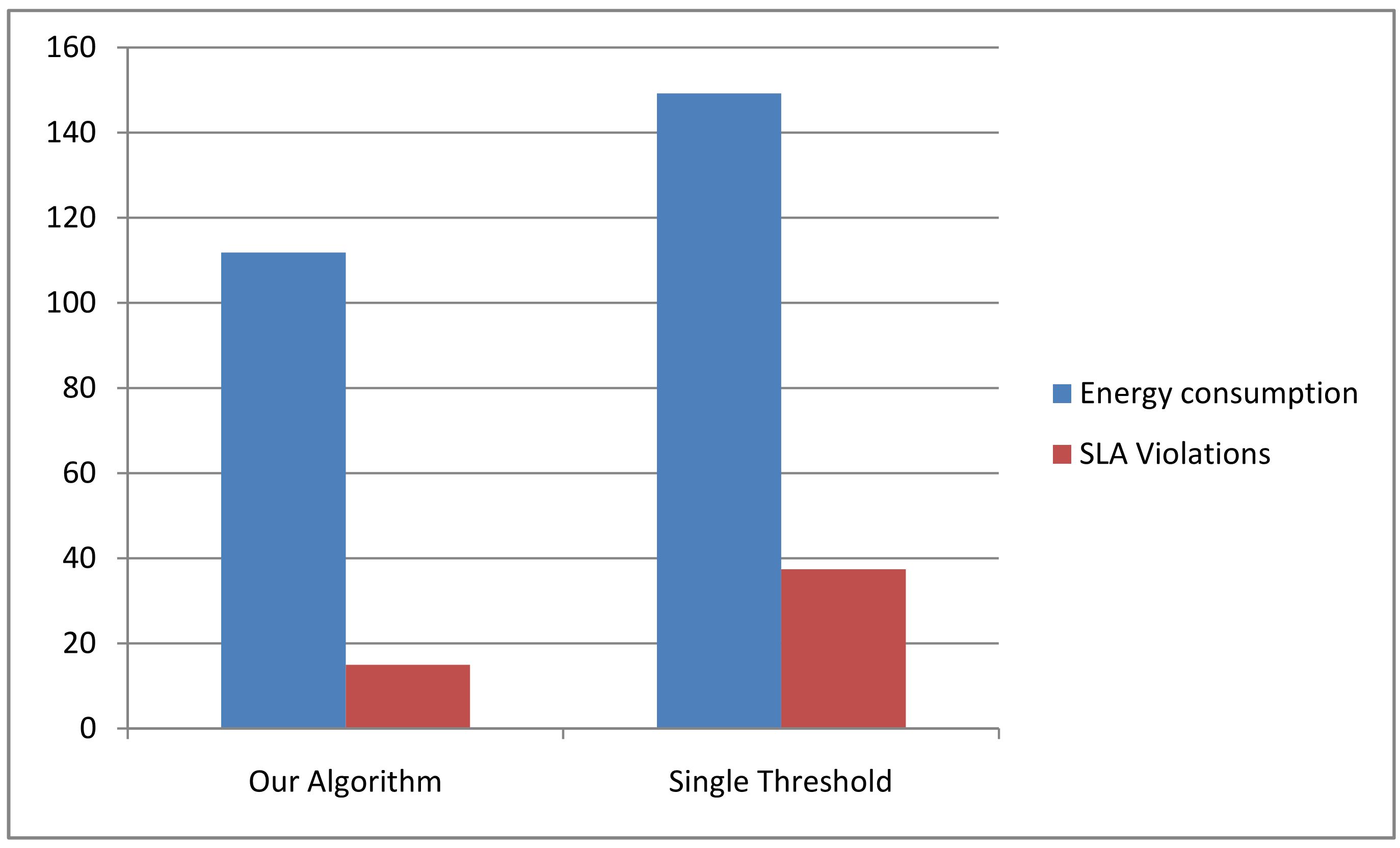}
 \caption{The graph demonstrates the effectiveness of our algorithm against
Single Threshold algorithm in terms of both energy consumption (in kWh) and also
number of SLA violations.}
 \label{comparison_with_ST}
\end{figure}

We have conducted several experiments with various workloads on both Single
Threshold and our algorithm. We have chosen the best configuration for our
algorithm, i.e., with $U_{up}$ = 0.75, $U_{down}$ = 0.15, \emph{buffer} = 0.15,
Method 2, \emph{Similarity Threshold} = 0.6. And for Single Threshold algorithm a
threshold of 0.75 is used. In Figure \ref{comparison_with_ST}, we see a
considerable amount of energy savings with our algorithm, saving up to 21\% of
energy. While in terms of number of SLA violations, our algorithm performs very
well maintaining up to 60\% more SLA guarantees.

\section{Conclusion and Future Work}
\label{sec-conclusion}

In this paper, we proposed algorithms that try to conserve energy in cloud data
centers. We discussed the Allocation Algorithm which tries to consolidate the
virtual machines on physical machines taking into consideration of resource
usage characteristics. The similarity model that is discussed tries to avoid SLA
violations and allocates virtual machines accordingly. The Scale-up and
Scale-down algorithms keep track of resource usage of each physical machine and
dynamically rebalance the data center based on the utilization. We have
successfully evaluated our algorithms against Single Threshold algorithm and
they show a considerable amount of energy savings.

Future directions to this work include predicting the resource usage patterns of
virtual machines and take dynamic rebalancing decisions ahead of threshold
condition. We would like to analyze the efficacy of Machine Learning algorithms
in training of scheduling and rebalancing decisions of the data center. 

In our algorithms, we have put the physical machines to a \emph{standby} mode
when they are not in use. We would also want to analyze the effect of our
algorithms if the machines which are not used are \emph{switched off}. Switching
off the machines would have a significant improvement in energy savings. But at
the same time, there is a need to analyze the effect of time delay in switching
on the machines, which the scheduler has to anticipate during its scheduling
decision.

\bibliographystyle{abbrv}
\bibliography{/Users/GreatGod/Research/My/mybibliography}

\begin{thebibliography}{10}

\bibitem{1e}
1E.
\newblock {Server Energy and Efficiency Report}.
\newblock Technical report, 1E, 2009.

\bibitem{energy-prop-comp}
L.~Barroso and U.~Holzle.
\newblock The case for energy-proportional computing.
\newblock {\em Computer}, 40(12):33 --37, dec. 2007.

\bibitem{SingleThreshold}
A.~Beloglazov and R.~Buyya.
\newblock Energy efficient allocation of virtual machines in cloud data
  centers.
\newblock In {\em Proceedings of the 2010 10th IEEE/ACM International
  Conference on Cluster, Cloud and Grid Computing}, CCGRID '10, pages 577--578,
  Washington, DC, USA, 2010. IEEE Computer Society.

\bibitem{buyya-vision}
R.~Buyya, C.~S. Yeo, S.~Venugopal, J.~Broberg, and I.~Brandic.
\newblock Cloud computing and emerging it platforms: Vision, hype, and reality
  for delivering computing as the 5th utility.
\newblock {\em Future Gener. Comput. Syst.}, 25(6):599--616, June 2009.

\bibitem{douglis}
F.~Douglis and J.~Ousterhout.
\newblock Transparent process migration: Design alternatives and the sprite
  implementation.
\newblock {\em Software - Practice and Experience}, 21:757--785, 1991.

\bibitem{pinheiroxx}
{E. Pinheiro, R. Bianchini, E. V. Carrera and T. Heath}.
\newblock Dynamic cluster reconfiguration for power and performance.
\newblock In {\em Power and energy management for server systems}. IEEE
  Computer Society, November 2004.

\bibitem{epa}
EPA.
\newblock {EPA Report to Congress on Server and Data Center Energy Efficiency}.
\newblock Technical report, U.S. Environmental Protection Agency, 2007.

\bibitem{GreenPeace}
{Gary Cook, Jodie Van Horn}.
\newblock {How dirty is your data? A Look at the Energy Choices That Power
  Cloud Computing}.
\newblock Technical report, Green Peace International, April 20, 2011.

\bibitem{hsu-hpc}
C.~Hsu and W.~Feng.
\newblock A power-aware run-time system for high-performance computing.
\newblock In {\em SC '05: Proceedings of the 2005 ACM/IEEE conference on
  Supercomputing}, page~1, Washington, DC, USA, 2005. IEEE Computer Society.

\bibitem{khalidi}
Y.~A. Khalidi, J.~M. Bernabeu, V.~Matena, K.~Shirriff, and M.~Thadani.
\newblock Solaris mc: a multi computer os.
\newblock In {\em Proceedings of the 1996 annual conference on USENIX Annual
  Technical Conference}, pages 16--16, Berkeley, CA, USA, 1996. USENIX
  Association.

\bibitem{buyya-dvs}
K.~H. Kim, R.~Buyya, and J.~Kim.
\newblock Power aware scheduling of bag-of-tasks applications with deadline
  constraints on dvs-enabled clusters.
\newblock In {\em Proceedings of the Seventh IEEE International Symposium on
  Cluster Computing and the Grid}, CCGRID '07, pages 541--548, Washington, DC,
  USA, 2007. IEEE Computer Society.

\bibitem{lee-dvs}
Y.~C. Lee and A.~Y. Zomaya.
\newblock Minimizing energy consumption for precedence-constrained applications
  using dynamic voltage scaling.
\newblock In {\em Proceedings of the 2009 9th IEEE/ACM International Symposium
  on Cluster Computing and the Grid}, CCGRID '09, pages 92--99, Washington, DC,
  USA, 2009. IEEE Computer Society.

\bibitem{DRR}
C.-C. Lin, P.~Liu, and J.-J. Wu.
\newblock Energy-aware virtual machine dynamic provision and scheduling for
  cloud computing.
\newblock In {\em Cloud Computing (CLOUD), 2011 IEEE International Conference
  on}, pages 736--737, july 2011.

\bibitem{nitesh}
N.~Maheshwari, R.~Nanduri, and V.~Varma.
\newblock Dynamic energy efficient data placement and cluster reconfiguration
  algorithm for mapreduce framework.
\newblock {\em Future Generation Comp. Syst.}, 28(1):119--127, 2012.

\bibitem{perez-reconfig}
M.~S. P\'{e}rez, A.~S\'{a}nchez, J.~M. Pena, and V.~Robles.
\newblock A new formalism for dynamic reconfiguration of data servers in a
  cluster.
\newblock {\em Journal of Parallel and Distributed Computing}, 65(10):1134 --
  1145, 2005.
\newblock Design and Performance of Networks for Super-, Cluster-, and
  Grid-Computing Part I.

\bibitem{glunix}
D.~Petrou, S.~H. Rodrigues, A.~Vahdat, and T.~E. Anderson.
\newblock Glunix: A global layer unix for a network of workstations.
\newblock {\em Softw., Pract. Exper.}, 28(9):929--961, 1998.

\bibitem{pinheiro}
E.~Pinheiro, R.~Bianchini, E.~V. Carrera, and T.~Heath.
\newblock Load balancing and unbalancing for power and performance in
  cluster-based systems.
\newblock In {\em In Workshop on Compilers and Operating Systems for Low
  Power}, 2001.

\bibitem{virtualization}
J.~E. Smith and R.~Nair.
\newblock The architecture of virtual machines.
\newblock {\em Computer}, 38:32--38, May 2005.

\end{thebibliography}

\end{document}